\documentclass[checkin,showpacs,aps,prb]{revtex4-1}

\newcommand{\ub}{\mu_{\rm B}}
\newcommand{\bn}{\begin{enumerate}}
\newcommand{\en}{\end{enumerate}}
\newcommand{\ba}{\begin{eqnarray}}
\newcommand{\ea}{\end{eqnarray}}

\usepackage{graphicx,color}

\newcommand{\be}{\begin{equation}}
\newcommand{\ee}{\end{equation}}

\newcommand{\et}{{\it et al. }}
\newcommand{\ete}{{\it et al.}}

\def\prl{{ Phys. Rev. Lett. }}
\def\prx{{ Phys.\ Rev.\ X\ }}

\def\prb{{ Phys. Rev. B }}

\begin{document}

\newcommand{\clr}{\color{black}}












\title{Ultrafast reduction of
exchange splitting in ferromagnetic nickel
}

\author{G. P. Zhang}

 \affiliation{Department of Physics, Indiana State University,
   Terre Haute, IN 47809, USA }

\author{Y. H. Bai}

\affiliation{Office of Information Technology, Indiana State
  University, Terre Haute, IN 47809, USA }

\author{Thomas F. George}

\affiliation{Office of the Chancellor and Center for Nanoscience
  \\Departments of Chemistry \& Biochemistrony and Physics \& Astronomy
  \\University of Missouri-St. Louis, St.  Louis, MO 63121, USA }

\date{\today}

\begin{abstract}
{ A decade ago Rhie \et (Phys. Rev. Lett. {\bf 90}, 247201
  (2003)) reported that when ferromagnetic nickel is subject to an
  intense ultrashort laser pulse, its exchange splitting is reduced
  quickly.  But to simulate such reduction remains a big
  challenge. The popular rigid band approximation (RBA), where both
  the band structure and the exchange splitting are held fixed before
  and after laser excitation, is unsuitable for this purpose, while
  the time-dependent density functional theory could be
  time-consuming. To overcome these difficulties, we propose a
  time-dependent Liouville and density functional theory (TDLDFT) that
  integrates the time-dependent Liouville equation into the density
  functional theory. As a result, the excited charge density is
  reiterated back into the Kohn-Sham equation, and the band structure
  is allowed to change dynamically.  Even with the ground-state
  density functional, a larger demagnetization than RBA is found;
  after we expand Ortenzi's spin scaling method into an
  excited-state (laser) density functional, we find that the exchange
  splitting is indeed strongly reduced, as seen in the experiment.
  Both the majority and minority bands are shifted toward the Fermi
  level, but the majority shifts a lot more.  The ultrafast reduction
  in exchange splitting occurs concomitantly with
  demagnetization. While our current theory is still unable to yield
  the same percentage loss in the spin moment as observed in the
  experiment, it predicts a correct trend that agrees with the
  experiments.  With a better functional, we believe that our results
  can be further improved.}
\end{abstract}




 \maketitle

\section{Introduction}

Laser-induced ultrafast demagnetization
\cite{eric,ourreview,rasingreview} presents a new opportunity for
magnetic storage technology, as it significantly shortens the
read/write time,\cite{stanciu} a necessity for large data storage
devices. This has attracted extensive investigations both
theoretically \cite{prl00,bk,georgejmmm,carva2013,krauss,
  jap08,prb09,lefkidis09,ch,tows,es, illg, bar,super,
  krieger,ajs2013,guidoni,vahaplar} and
experimentally.\cite{esc,rudo,wie,sand,mathias,mangin,moi,sch,tur} The
method of choice to investigate such a fast demagnetization is the
time-resolved magneto-optical Kerr effect (TRMOKE). Despite earlier
debates\cite{koopmansprl,guidoni} on the suitability of TRMOKE for
femtomagnetism,\cite{ourreview} it is now generally agreed that by
carefully removing the nonmagnetic contribution from the Kerr
ellipticity and rotation signal, one can access the spin moment change
in the time domain.  Such a connection has been established by
comparing and contrasting the optical and magnetic response functions
using the first-principles method.\cite{np09,prb12} Recently, an
analytic relation has been found between the spin angular momentum and
the off-diagonal susceptibility.\cite{epl15} Another technique which
complements TRMOKE is the time- and spin-resolved photoemission
(TSRPE). It is capable of resolving the spin momentum change in the
crystal momentum space.\cite{mingsu14} In TSRPE, a laser pulse first
excites electrons from the spin-polarized valence band to the
conduction band, and a second pulse ionizes the electrons to the
vacuum. Since the energy of the emitted spin-polarized electron
reflects its original valence band energy, the exchange energy
splitting can be monitored.  However, whether the exchange splitting
is collapsed has been controversial.  The first TSRPE was reported a
long time ago,\cite{scholl} but the result has not been reproduced.
More recent experimental studies are extended to Gd.  Carley \et
\cite{carley} investigated the exchange-split $\Sigma$ valence bands
of gadolinium and found that the majority and minority bands both move
closer to the Fermi surface, but differ on the time scale.  In a
ferromagnetic nickel thin film, Rhie \et \cite{rhie} demonstrated the
collapse of the magnetic exchange splitting.  Pickel \et \cite{pickel}
used photoemission to identify the spin-orbit hybridization points in fcc
Co. Weber \et \cite{weber} tried to compare TRMOKE with TSRPE, but
their results were not conclusive since different experimental
conditions were used for TRMOKE and TSRPE.  In Fe, Carpene \et
\cite{carpene2015} showed recently that the modification of the
electronic band structure upon laser excitation is small, but argued
that in nickel the collapse of the exchange splitting might be
justified.

Theoretically, several different approaches are available. One is
based on the rigid band approximation (RBA),\cite{har,kir,hau,lee}
which has been used in semiconductors\cite{koch,lingos} as well as
ferromagnets.\cite{prl00,jap08,tows} Under RBA, the band structure is
not allowed to change before and after laser excitation, and thus it
fixes the exchange splitting.  For this reason, RBA is unsuitable for
the exchange splitting. It has been argued that the rigid band
structure calculations will never be in quantitative agreement with
experiments, irrespective of the investigated microscopic scattering
mechanism.  \cite{ajs2013} In a simplified model calculation, Mueller
\et \cite{mueller} proposed a scheme that allows the charge density to
dynamically affect the exchange splitting change, which they call the
feed-back effect.  More recently, the time-dependent density
functional theory (TDDFT) was employed to investigate the
demagnetization,\cite{krieger} but it is very time-consuming to carry
out such a calculation. This motivated Wang \et to develop a two-step
time propagation.\cite{wang}



%


\newcommand{\ik}{i{\bf k}} \newcommand{\jk}{j{\bf k}}

\newcommand{\lk}{l{\bf k}}



\newcommand{\br}{{\bf r}}

In this paper, we propose an alternative scheme to TDDFT, which is
less time-consuming. We merge the standard density functional theory
(DFT) with the time-dependent Liouville equation, so the excited-state
density is reiterated back into the Kohn-Sham (KS) equation, thus
going beyond the rigid band approximation.  The self-consistent
calculation converges the KS orbitals on the excited-state potential
surface, different from the ground-state calculation. We call this
scheme the time-dependent Liouville density functional theory, or
TDLDFT.  We find that even with the ground-state functional, the
demagnetization is an order of magnitude larger than RBA.  Since most
of the existing exchange correlation functionals are geared toward the
ground-state properties, to properly describe the excited-state
property, we implement a functional based on the Ortenzi spin scaling
function, where the spin polarization acts upon the system
self-consistently.  We find that both the majority and minority bands
are shifted toward the Fermi surface, but the amount of shift is
different. The majority band moves upward by 0.26 eV, while the
minority one moves downward by only 0.03 eV. As a result, the exchange
splitting is reduced.  Interestingly, we find that the exchange
splitting change correlates well with the spin moment change in
ferromagnetic nickel. We have tested three functionals, with the
largest spin moment reduction reaching 10\%. Although this is still
below the experimental value, the trend seems promising. With a better
functional, we expect that our results can be systematically improved
upon.

 { The rest of the paper is arranged as follows. In Sec. II, we
   present the theoretical formalism. Section III is devoted to the
   results and discussions on (i) the comparison between the
   rigid-band approximation and the TDLDFT calculation, (ii) the
   development of the excited-state (laser) functionals, and (iii) the
   collapse of the exchange splitting. The paper is concluded in
   Sec. IV. }

\section{Time-dependent Liouville density functional theory (TDLDFT)}

Our new idea comes from an important observation.  After a laser pulse
impinges on a 3$d$ magnet, a few electrons are excited out of the
Fermi sea, with 3$d$ holes left behind.  It is important to realize
that losing a few electrons around the Fermi level will significantly
weaken the exchange correlation,\cite{jpcm14,jpcm15,mentink,mueller}
creating a new potential for the entire system and setting off an
avalanche of spin change.  In the many-body picture,\cite{li} the
electron dynamics takes place on an excited-state potential surface
that can be very different from the ground-state one.  In the density
functional theory, the many-body correlation effect is captured
through the exchange-correlation functional, but now one has to solve
the Kohn-Sham equation self-consistently, so {\it the new potential
  must act upon itself}. We tested this idea of a static version of
this method in a prior study,\cite{jpcm15} where we saw a big effect
on the spin moment.

In our new algorithm, we first solve the Kohn-Sham
equation for the ground state, \be
\left [-\frac{\hbar^2\nabla^2}{2m_e}+v_{eff}^\sigma(\br) \right ]
\psi_{\ik}^\sigma(\br)=E_{\ik}^\sigma \psi_{\ik}^\sigma (\br).
\label{ks}
\ee Here the first term on the left-hand side is the kinetic energy,
$ \psi_{\ik}^\sigma (\br)$ and $E_{\ik}^\sigma$ are respectively the
eigenstate and eigenenergy of band $i$ and ${\bf k}$ point with spin
$\sigma$, and $v_{eff}^\sigma$ is determined by \be
v_{eff}^\sigma({\bf r})=v^\sigma({\bf r})+\int \frac{\rho^\sigma({\bf
    r}')}{|{\bf r}-{\bf r}'|} d{\bf r}'+v_{xc}^\sigma({\bf r}), \ee where
$v_{xc}^\sigma({\bf r})$ is the exchange-correlation potential,
$v_{xc}^\sigma({\bf r})=\delta E_{xc}[\rho^\sigma]/\delta \rho({\bf
  r})$.  We use the generalized gradient approximation for the
exchange-correlation energy functional.  The spin-orbit coupling is
included through the second-variational method,\cite{wien}  so the
following wavefunctions and eigenvalues have no spin index.

The laser excitation is computed by the time-dependent Liouville
equation,\cite{np09,prb12} \be i\hbar\frac{\partial \rho_{{\bf k},
    ij}}{\partial t}=[H_0+H_I, \rho_{{\bf k}, ij}] \label{liou},\ee
where $\rho_{{\bf k},ij}$ is the density matrix element between band
states $i$ and $j$ at the ${\bf k}$ point, $H_I$ is the interaction
between the laser and the system (see below for details), and $H_0$ is
the unperturbed Hamiltonian, $H_0=\sum_{i{\bf k}} E_{i{\bf k}}
|\psi_{i{\bf k}}\rangle\langle \psi_{i{\bf k}}|$. Different from the
rigid-band approximation (RBA), we only integrate a small time step,
$\Delta t$, typically one-eighth of the laser period, but within
$\Delta t$, we solve Eq. (\ref{liou}) accurately with a high tolerance
of $5\times 10^{-14}$. This method is identical to that of Wang \et
\cite{wang} who separate a single time step into two time steps by
expanding the real time wavefunction in terms of the adiabatic
eigenstate $\psi_{i{\bf k}}({\bf r},t)$ of the Hamiltonian at a
specific time step, while $\psi_{i{\bf k}}({\bf r},t)$ is only
approximately propagated.  Here we do not extrapolate between two time
steps since the spin excitation is much slower.

In the time-dependent density functional theory (TDDFT),\cite{tddft}
the time-dependent Kohn-Sham equation is solved in real time with a
very tiny time step since the time step is directly linked to the
inverse of the total energy, so TDDFT is often limited to an extremely
short time scale.  The new density is computed by \be \rho({\bf
  r},t)=\sum_{i{\bf k}}n_{i{\bf k}}\psi_{i{\bf k}}^*({\bf
  r},t)\psi_{i{\bf k}}({\bf r},t), \ee where $n_{i{\bf k}}$ is the
occupation number and is fixed in time from the beginning. In TDLDFT,
\be \rho({\bf r},t)=\sum_{i{\bf k}}\rho_{{\bf k};ii}(t)\psi_{i{\bf
    k}}^*({\bf r},t)\psi_{i{\bf k}}({\bf r},t), \ee where $\rho_{{\bf
    k};ii}$ is computed from the Liouville equation
(Eq. (\ref{liou})). In TDLDFT $\psi_{i{\bf k}}({\bf r},t)$ is the
adiabatic eigenstate at time $t$ and is not solved from the
time-dependent Kohn-Sham equation, thus saving lots of time, and is
useful for long-time dynamics that is actually observed
experimentally, in contrast to TDDFT.  The time step size is
determined by Eq. (\ref{liou}).  As one may realize from
Eq. (\ref{liou}), the Liouville equation gives the density matrix, not
the density itself. But since the premise of the density functional
theory is that the exchange-correlation potential is a functional of
the density, not the density matrix, when we assemble the density
$\rho({\bf r},t)$, the off-diagonal density matrix elements of
$\rho_{{\bf k};ij}$ are discarded. This points out a possible
extension of our current formalism in the future.

\begin{figure}
\includegraphics[angle=0,width=16cm]{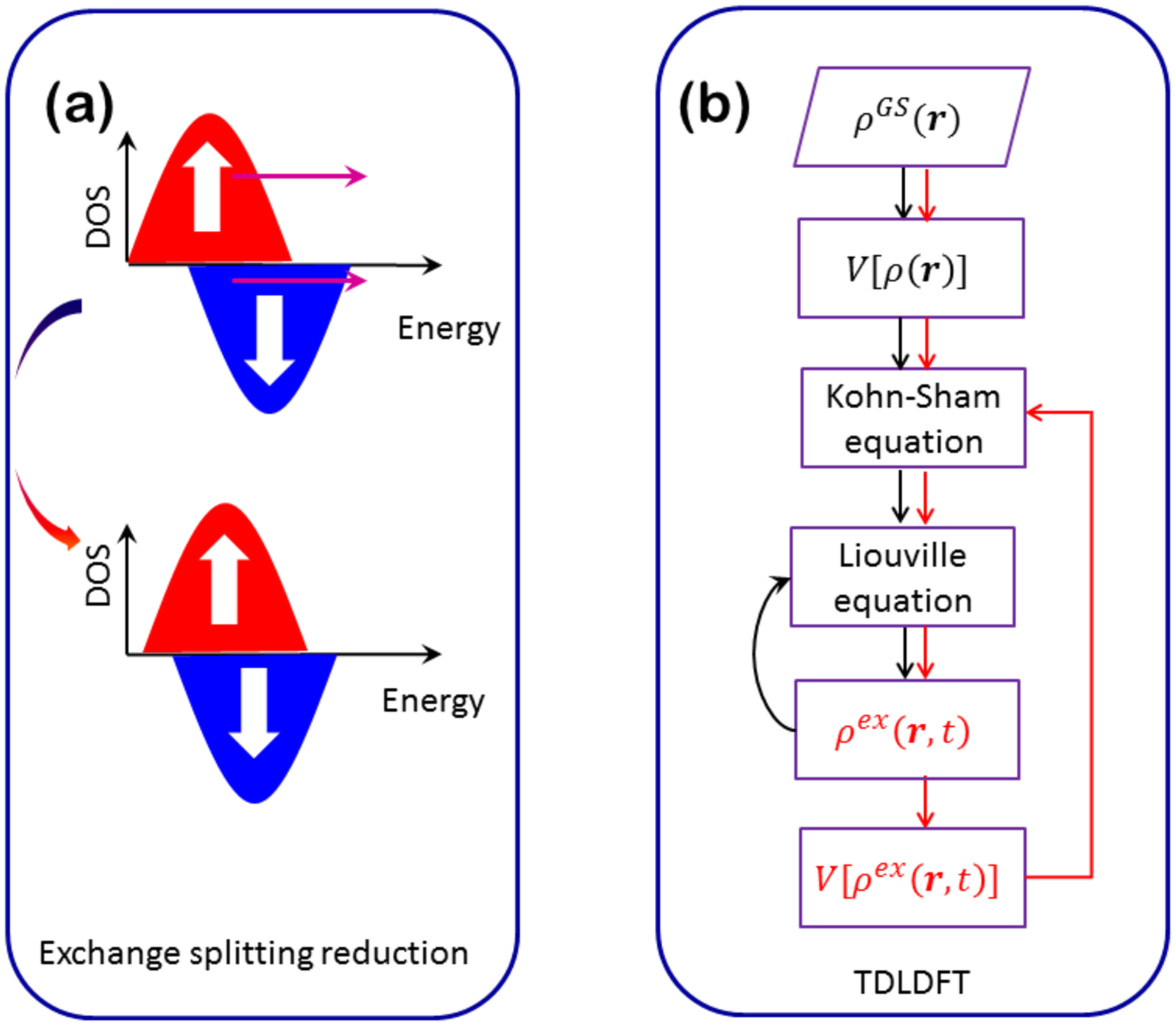}
\caption{ (a) Ultrafast laser-induced exchange splitting reduction.
  The laser pulse excites electrons out of the Fermi sea and weakens
  the exchange correlation. The minority and majority bands start to
  shift toward the Fermi level.  (b) Flowchart of the time-dependent
  Liouville density functional theory (TDLDFT).  The black arrows
  refer to the rigid-band approximation, while the red denote
  our new TDLDFT.  }
\label{fig1}
\end{figure}

To catch the many-body excitation, we iterate the resultant density
$\rho({\bf r}, t)$ back into the Kohn-Sham equation (\ref{ks}) and
solve it self-consistently under this excited density and thus the
time-dependent potential $v_{eff}({\bf r}, t)$.  Such a
self-consistent calculation is crucial since it essentially allows the
excited density to affect upon the system itself and thus catches
majority of the missing electron correlation and many-body effects in
RBA.  Figure \ref{fig1}(b) compares our TDLDFT algorithm (see the
flowchart with red arrows) with the RBA one (see the flowchart with
black arrows).  The TDLDFT employs an idea similar to a prior study by
Mueller \ete,\cite{mueller} who only implemented it for a model
system, but there are several major differences. Our method is
implemented at the first-principles level. We do not shift the bands
manually; instead we include the spin-orbit coupling to allow the spin
change. In comparison with TDDFT, the Liouville formalism-based TDLDFT
has another advantage. It naturally respects the Pauli exclusion
principle, which is extremely important for the system with many
electrons at a single ${\bf k}$ point. Once excited by laser pulses,
the occupation can be dynamically changed, without fixing the
occupation for the entire dynamics, which is closer to real dynamics.

\section{Results and discussions}

To demonstrate the power of TDLDFT, we take fcc Ni as an example.
Different from prior studies,\cite{krieger} our laser parameters are
very close to the experimental ones.\cite{eric} The interaction
between the laser field and the system is\cite{jpcm11} \be
H_I=\frac{e}{m} {\bf p}\cdot {\bf A}(t),\ee where $-e$ is the electron
charge, $m$ is the electron mass, ${\bf p}$ is the momentum operator,
and the vector potential ${\bf A}(t)$ is along the $z$ axis with a
Gaussian shape $ | {\bf A}(t) |=A_0 \exp(-t^2/\tau^2)\cos (\omega t)$,
with $A_0=0.03~\rm V fs/\AA$.  The duration is $\tau=60$ fs and the
photon energy $\hbar\omega$ is 2 eV, corresponding to the experimental
wavelength of 620 nm.\cite{eric} We note that the TDDFT
study \cite{krieger} used the three photon energies, i.e., 1.35, 2.73
and 5.42 eV, which do not match any experimental one.  Since TDDFT is
very time-consuming, an extremely short pulse was used.  For the same
reason, the number of ${\bf k}$ points was only ($8\times8\times 8$),
too few to converge the results.\cite{jap99} This makes the TDDFT results
difficult to compare with the experimental ones.  In our study, we use
a {\bf k} mesh of $(30 \times 30 \times 30)$, and we test its
convergence using a larger number of ${\bf k}$ points.  The transition
matrix element for the momentum operator is directly computed using
WIEN2k's optic code.\cite{claudia}

\subsection{Comparison between the rigid-band approximation and
  TDLDFT calculation}

To have a quantitative understanding of the spin moment reduction,
Figure \ref{fig2}(a) compares the RBA and TDLDFT spin moments as a
function of time.\cite{np09} Under RBA (long dashed line), the spin
moment reduces quickly from 0.637 $\ub$ to the minimum of 0.627 $\ub$
around 0 fs, but soon recovers to 0.634 $\ub$, or 0.47\%, consistent
with our prior study \cite{jap08} and also others.\cite{ajs2013} In
the following, we define the time at the spin moment minimum as the
demagnetization time $\tau_M$.  Such a sharp reduction and quick
recovery is the hallmark of the system overheating, where the
electrons are temporally held by the laser field in the excited states
(electrons are field-dressed), and they can not pass the excessive
energy to other unexcited electrons beyond the parent ${\bf k}$ point
within a single-particle picture.  \cite{mingsu14} Once the laser is
gone, only a few electrons are left in the excited states and majority
of the excited electrons return to low-energy states and the spin
moment is restored. It is clear that within RBA, the spin moment
reduction is much smaller than the experimental observation, but the
reason is simple.  Any transitions among band states \cite{mingsu14}
must obey the dipole selection rule; and any strong transitions must
have a large transition matrix element, and their transition energy
should match or be close to the photon energy of the incident
light. However, in solids, only a small number of ${\bf k}$ points
satisfy these conditions,\cite{mingsu14} which imposes a severe
constraint on any theory.  The superdiffusion model \cite{super} has a
larger spin moment reduction since the above conditions are abandoned,
as verified in an earlier study.\cite{jpcm15} In summary, RBA is a
single-shot non-self-consistent calculation and misses the dynamic
many-body effect on the system itself. Therefore, RBA fails to induce
a strong demagnetization.

The situation is quite different for TDLDFT. The solid line in Fig
\ref{fig2}(a) shows that the spin moment computed with TDLDFT is
reduced to 0.6149 $\ub$, or 3.5\%, nearly an order of magnitude larger
than the RBA calculation. Note that both RBA and TDLDFT use the same
laser parameters.  Such a reduction is robust, regardless of the generalized gradient approximation (GGA)
or local density approximation (LDA) used for the functional. Different from the RBA results, the spin
moment minimum is no longer at 0 fs, but instead shifts to $\tau_M=70$
fs; and the spin does not recover within 100 fs either, fully
consistent with the experimental observation.\cite{eric} This is
encouraging. We wonder whether TDLDFT can explain how the laser
amplitude affects $\tau_M$.

Experimentally it is well known \cite{bk,carpene,chan} that $\tau_M$
becomes shorter with a weak laser, but theoretically, $\tau_M$ is
nearly independent of the laser field amplitude within the rigid-band
approximation, a finding that is often used as evidence for phonon
involvement,\cite{bk} which further complicates the issue. We consider
two laser amplitudes, 0.01 and 0.03 $\rm Vfs/\AA$.  Figure
\ref{fig2}(a) (see the vertical bars) shows that as we decrease the
field vector potential from 0.03 to 0.01 $\rm Vfs/\AA$, while keeping
the rest of parameters fixed, $\tau_M$ indeed reduces from 70 to 45
fs. There is no need to invoke the phonon contribution.  The reason
for this dependence is straightforward. For a weaker laser, only those
transition states with the transition energy matching the photon
energy are strongly excited; as a result, their response is impulsive
and faster. When the laser becomes stronger, the low-lying states
close to the Fermi surface start to contribute, so the demagnetization
slows down.

\subsection{Functional for the excited states}

While our results are encouraging, quantitatively our spin moment
reduction is still lower than the experimental data.  Krieger \et
\cite{krieger} did observe a much larger reduction, but with a laser
amplitude at least two orders of magnitude higher than the
experimental one.  We want to understand anything missing from our
theory.  As discussed above, within DFT, many-body effects are
included through the exchange-correlation functional, but all the
density functionals in use are highly geared toward the ground-state
properties; and there is no well-established functional for excited
states if it exists.  GGA strongly favors a magnetic solution in the
ground state; DFT gives too high Curie temperatures for all the
transition metals.\cite{kubler} A common practice to overcome this
problem is to compute the effective exchange interaction which gives a
much better Curie temperature.

Ortenzi \et \cite{ortenzi} suggested a different approach to rescale
the exchange and correlation potential for the spin part, while
keeping the charge potential fixed. However, the Ortenzi formalism is
completely static.  To describe the laser-induced ultrafast
demagnetization, we develop their method into a time-dependence
functional, with the spin-polarized potential \ba
V_{\uparrow}^{new}({\bf r},t)&=&\frac{1}{2}\left
((1+f(t))V_\uparrow({\bf r},t)+(1-f(t))V_{\downarrow}({\bf r},t)
\right ), \\ V_{\downarrow}^{new}({\bf r},t)&=&\frac{1}{2}\left
((1+f(t))V_\downarrow({\bf r},t)+(1-f(t))V_{\uparrow}({\bf r},t)
\right ), \ea where $V_{\uparrow(\downarrow)}$ is the potential for
spin up (down).  $f(t)$ is the time-dependent spin scaling and is a
functional of the spin density scaling factor $\xi$ for the Stoner
kernel.  $\xi$ is fit to the magnetic moment change with pressure
($\xi=0.88$ in their case).\cite{ortenzi} Here we make two
extensions. First, we choose $f(t)=\xi^\alpha$ (other forms are
presented below); Ortenzi's functional is recovered if
$\alpha=1$. Second, we redefine $\xi$ as the ratio of the spin moment
to the initial value or $\xi(t)=M_z(t)/M_z(-\infty)$, so the
time-dependent exchange-correlation potential is self-consistently
rescaled by the spin moment change. Thus, in our formalism $\xi$ is no
longer a fitting parameter. Instead, it has a physical meaning as it
attenuates the strength of the exchange-correlation potential to
reflect the diffusive nature of the excited states \cite{jpcm14} and
builds in a memory effect.\cite{tddft} This second step allows us to
smoothly connect the ground-state functional, where $\xi(t)=1$, to the
excited-state functional, while keeping intact all the good features
of density functionals in GGA or LDA.

\begin{figure}
\includegraphics[angle=270,width=16cm]{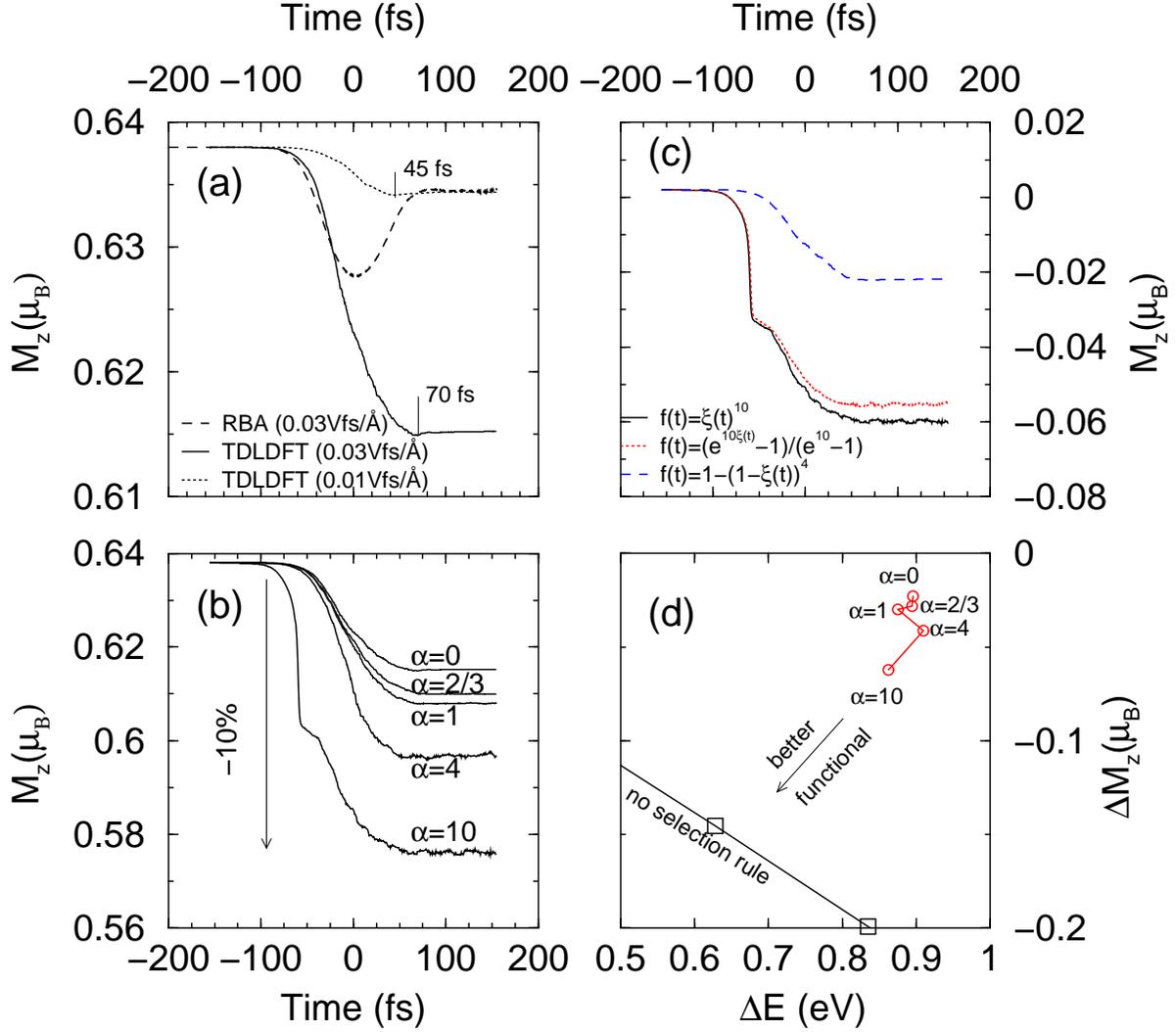}
\caption{(a) Comparison of the spin moment reduction between the rigid-band approximation (dashed line, RBA) and the TDLDFT
  calculation (solid line, TDLDFT) as a function of time. The dotted line
  denotes the spin change when the laser vector potential is reduced
  to $A_0=0.01 \rm V fs /\AA$. (b) Spin moment change as a function
  of the spin attenuation factor $\alpha$.  With the largest $\alpha$,
  we achieve a 10\% reduction. (c) Influence of the functionals on the
  spin change. The dotted line refers to the exponential functional,
  while the long-dashed one the power functional. (d) Spin moment
  change versus the absorbed energy. We expect that a better agreement
  with the experiment will
  be reached if we find a better excited (laser) functional. The empty
  boxes denote the prior results,\cite{jpcm15}  without taking into
  account the selection rule.
}\label{fig2}
\end{figure}

Our algorithm gets the best of both worlds: From the density
functional theory, we effectively avoid the many-body problem and on
the excited potential get excited-state properties, while from
Liouville dynamic formalism, we add ``time'' to the original static
DFT. This avoids the limitation of the tiny time step in TDDFT,
\cite{tddft}  unphysical absorption peak shifting,\cite{provorse}  and
the time-dependent response frequencies of the Kohn-Sham response
function even in the absence of an external field.\cite{fuks}   It can be
easily incorporated into all the existing codes. In our study, we have
implemented this algorithm in the Wien2k code.\cite{wien}   $\alpha$
is used to control the level of attenuation on the spin, which will be
called the spin attenuation factor below. The first line in Fig.
\ref{fig2}(b) is our  data with $\alpha=0$ (same as the
solid line in Fig. \ref{fig2}(a)). We gradually enhance the spin
attenuation factor $\alpha$, {\it while keeping the rest of the parameters
  unchanged}, and we find that the amount of spin reduction increases
sharply. The main shape of the spin moment reduction does not change much from $\alpha=0$ to 4. When we increase $\alpha$ to 10, we find
that the spin reduction reaches -10\%, much closer to the experimental
results.\cite{eric}   This demonstrates the great potential of the
density functional theory as an enabling theory to describe the strong
demagnetization. Interestingly, at $\alpha=10$, $M_z(t)$ shows a
kink around -55 fs, before the final minimum is reached.

 To better understand the demagnetization, we test two additional
 functionals, \ba
 &f_1(t)&=\frac{e^{10\xi(t)}-1}{e^{10}-1},\\ &f_2(t)&=1-(1-\xi(t))^4,
 \ea where $f_1(t)$ has an exponential dependence and $f_2(t)$ has a
 power dependence.  Figure \ref{fig2}(c) shows the net reduction of
 the spin moment.  The kink on $M_z(t)$ is directly connected to the
 highly nonlinear dependence of the functional on $\xi(t)$, thus
 appearing in both functionals.  The power functional $f_2(t)$ has no
 such kink, since its nonlinearity is smaller than the other two. This
 points to a promising new frontier by developing new functionals for
 the laser excitation, or laser functional. To quantify how far we are
 away from the best functional, Fig. \ref{fig2}(d) shows the spin
 moment reduction versus the absorbed energy (the total energy
 difference $\Delta E=E_{final}-E_{initial}$) for each $\alpha$. This
 is the absolute measure on the absolute energy and spin moment
 scale. Our target is the demagnetization line obtained by our prior
 study \cite{jpcm15} without considering the optical selection rule
 (see the line with the empty boxes in the lower left corner in
 Fig. \ref{fig2}(d)), which agrees with the experimental result if we
 assume 12.5\% absorption efficiency.\cite{jpcm15} Note that our
 present energy convergence criterion is much smaller than the energy
 change reported in the figure.  When $\alpha=0$, both $\Delta M_z$
 and $\Delta E$ are small. When we increase $\alpha$ to 2/3 and 1, we
 see that the system absorbs less energy, not more energy, but with a
 larger spin reduction, a single most important finding.  This
 demonstrates that our algorithm samples a much broader energy space,
 where the spin moment reduction does not need lots of
 energy. Naturally, this trend can not hold for any $\alpha$ and for
 any functional. When we increase $\alpha$ to 4, the usual trend is
 restored, i.e., the larger $\Delta E$, the larger $\Delta M_z$.  We
 strongly believe that if a better functional is found, a better
 agreement can be reached. In particular, we see that when we increase
 $\alpha$ to 10, our result is closer to the experimental one.

\subsection{Collapse of the exchange splitting}

\begin{figure}
\includegraphics[angle=270,width=12cm]{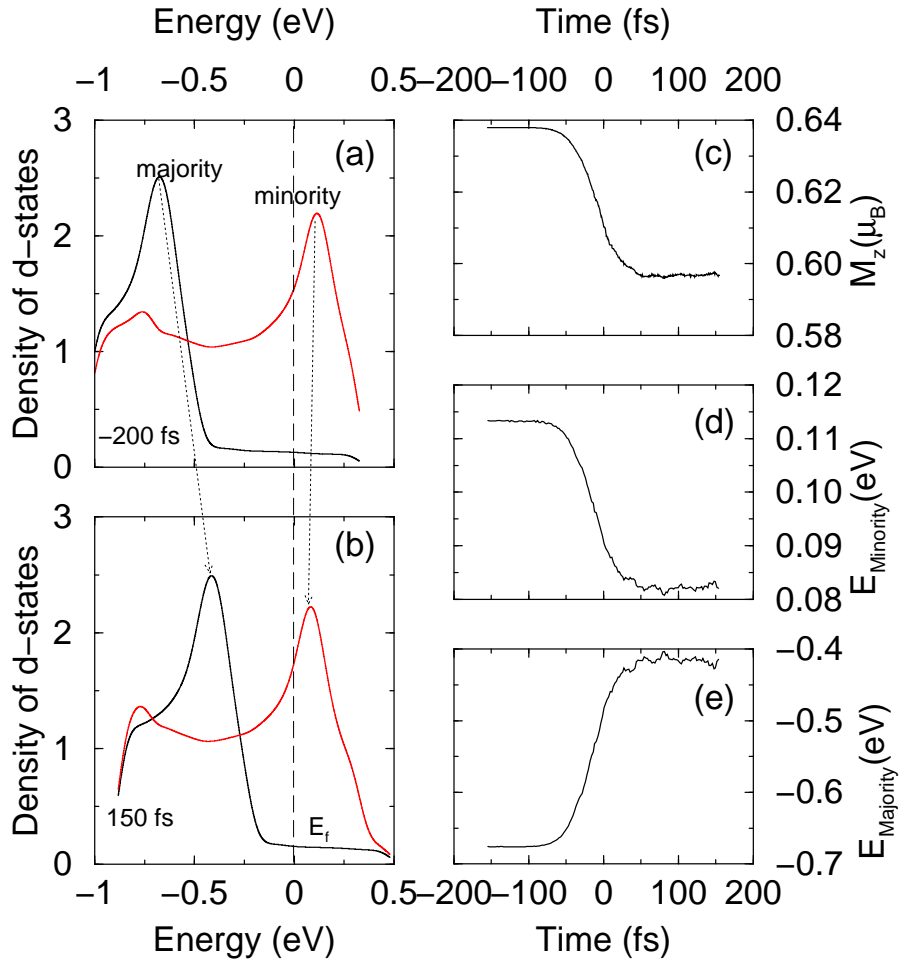}
\caption{Density of $3d$ states (a) before and (b) after the laser
  excitation. The majority and minority peaks are
  clearly shifted toward the Fermi level, which is set at 0 eV. (c)
  Spin moment change as a function of time for a spin attenuation factor of
  $\alpha=4$.  (d) and (e) Peak energy of the minority and majority
  bands as a function of time. The majority band shifts 0.26 eV, while
  the minority 0.03 eV.  }\label{fig3}
\end{figure}

The realization of the strong demagnetization opens the door to
understand the exchange splitting reduction. Figures \ref{fig3}(a) and
(b) show the density of $d$ states at -200 fs (in the absence of the
laser) and 150 fs (after the laser excitation). The Fermi level is set
at 0 eV. It is clear that both the majority and minority bands shift
toward the Fermi level, but the majority shifts more, a similar
finding reported for 4$f$ Gd.\cite{carley} As a result, the exchange
splitting is significantly reduced, in agreement with the prior
experimental results.\cite{rhie}

But the photoemission can not directly assign the exchange splitting
quenching to the demagnetization, since it probes only a small portion
of the Brillouin zone.\cite{weber}  This missing link is provided in
Figs. \ref{fig3}(c), (d) and (e). Figure \ref{fig3}(c) reproduces the
spin moment change for $\alpha=4$. The peak energies for the majority
and minority bands, $E_{majority}$ and $E_{minority}$, are shown in
Fig. \ref{fig3}(d) and Fig. \ref{fig3}(e), respectively. It is very
clear that the demagnetization follows $E_{majority}$ and
$E_{minority}$ closely.  This result represents the first theoretical
confirmation of a long speculation as to how the demagnetization and
exchange splitting are correlated in ferromagnetic Ni. Since different
materials differ a lot in terms of the electronic and magnetic
properties, further investigation is necessary for other
materials. Recently, Andres \et \cite{andres} found the spin mixing in
the surface state is not related to the exchange splitting change in
Gd.  Frietsch \et \cite{frietsch} suggested that the initial drop of
the exchange splitting also follows the magnetic moment change in 5$d$
electrons, not 4$f$ electrons, but since their Landau-Lifshitz-Gilbert
equation is module-conserved, they can only investigate the spin
precession, not a true demagnetization. Our study provides a much
needed theory.

\subsection{Beyond density functional}

The density functional theory represents a
state-of-the-art first-principles technique to investigate the
ground-state electronic and magnetic properties. To compute the
excited states, DFT has intrinsic difficulties. First of all, nearly
all the studies employ the ground-state functional to compute the
excited-state properties.  Second, in ferromagnets the spin wave
excitation is formally not included in DFT, since only the density
itself enters the theory.  Our spin attenuation factor $\alpha$ only
partially remedies this shortcoming. At $\alpha=0$, the agreement with
the experiment is poor, since the spin-polarized density is still not
enough to weaken the exchange interaction. A larger $\alpha$ leads to
a better agreement, since it reduces the exchange correlation more,
but too big a value leads to an unphysical kink as seen in
Fig. 2(b). A better functional is necessary.

To go beyond the density functional, one may include the relaxation of
the collective spin waves excitation.  However, there are many-body
interactions that are hard to treat.  In the traditional spin wave
theory, the spin moment reduction is due to the creation of magnons (spin wave quanta), often driven by a thermal field. The exchange splitting is certainly affected by the spin-wave relaxation. A weaker
spin wave will lead to a smaller energy difference between the spin-up
and spin-down electrons and a smaller spin moment overall.  The
difficulty is that the excitation by a femtosecond laser pulse is not
the region that the spin wave theory can handle easily, since the
interaction field is an electric field. One possible solution is to
incorporate the spin wave excitation idea into the existing density
functional theory, so the influence of the spin wave excitation on the
exchange splitting and demagnetization can be closely examined.
Clearly, additional research is needed along this direction.

\section{Conclusion}

We have developed an algorithm that integrates the time-dependent
Liouville scheme into the density functional theory. Our algorithm
respects the optical dipole selection rule. Importantly, this scheme
overcomes two major hurdles -- rigid-band approximation and
ground-state density functional -- and leads to a strong
demagnetization.  The key to our success is that we allow for the
excited exchange-correlation potential to act upon the system itself,
so those ${\bf k}$ points which are not optically accessible also
experience the change of the excited-state potential. We show that
without introducing an additional functional, a straightforward TDLDFT
calculation leads to a 3.5\% reduction, nearly an order of magnitude
higher than the rigid-band approximation results. Once we introduce
the Ortenzi functional, the maximum reduction within our current
functionals reaches 10\%. We expect that with a better functional, a
better agreement with the experiment can be reached in the future. As
a direct consequence of our current study, we can now directly
correlate the exchange splitting quenching to the demagnetization,
confirming the prior time- and spin-resolved photoemission
experiments. Since our method is relatively simpler than TDDFT, we may
be allowed to suggest that our method may find some applications to
laser-induced ultrafast dynamics in high-temperature superconductors
and other complex magnets. These systems have many more atoms in a
unit cell, so the TDDFT calculation might be extremely time consuming.
Another advantage that we notice is that it may include the lattice
vibration at a lower cost.  Research along this direction is underway.

\acknowledgments This work was solely supported by the U.S. Department
of Energy under Contract No. DE-FG02-06ER46304. Part of the work was
done on Indiana State University's quantum cluster and
high-performance computers.  The research used resources of the
National Energy Research Scientific Computing Center, which is
supported by the Office of Science of the U.S. Department of Energy
under Contract No. DE-AC02-05CH11231. This work was performed, in
part, at the Center for Integrated Nanotechnologies, an Office of
Science User Facility operated for the U.S. Department of Energy (DOE)
Office of Science by Los Alamos National Laboratory (Contract
DE-AC52-06NA25396) and Sandia National Laboratories (Contract
DE-AC04-94AL85000).




\clearpage

\end{document}